\documentclass [12pt]{article}
\usepackage{graphicx}

\begin{document}

{\bf States with identical steady dissipation rate: 
Role of kinetic constants in enzyme catalysis}

\vskip .1in

Kinshuk Banerjee and Kamal Bhattacharyya\footnote{Corresponding 
author; e-mail: pchemkb@gmail.com}

\vskip .05in
{\it Department of Chemistry, University of Calcutta, 
92 A.P.C. Road,

Kolkata 700 009, India.}

\begin{abstract}

A non-equilibrium steady state is characterized by a non-zero 
steady dissipation rate. Chemical reaction systems under suitable 
conditions may generate such states. We propose here 
a method that is able to distinguish states with 
identical values of the steady dissipation rate. This 
necessitates a study of the variation of the 
entropy production rate with the experimentally observable 
reaction rate in regions close to the steady states. 
As an exactly-solvable test case, we choose the problem 
of enzyme catalysis. 
Link of the total entropy production with the 
enzyme efficiency is also established, 
offering a desirable connection with the inherent 
irreversibility of the process. 
The chief outcomes are finally noted in a more general reaction 
network with numerical demonstrations. 

\end{abstract}

PACS: 05.70.Ln, 82.39.-k, 82.20.-w

\vskip .05in

Keywords: Entropy production rate, Dissipation, Enzyme efficiency,\\  Reaction network

\section{Introduction}

A major shift in the field of thermodynamics in the last 
century was from idealized equilibrium processes to 
natural irreversible processes [1-4].
Chemical reactions 
continue to play a pivotal role in this development 
and provide significant motivation in studying the non-equilibrium 
thermodynamic properties of systems {\it in vitro} 
as well as {\it in vivo} [5-10]. 
Since a closed system always tends to thermodynamic equilibrium (TE), 
a natural generalization in the theory of irreversible 
thermodynamics has been achieved via 
the concept of a steady state \cite{denb,Ono}. In this regard, 
the quantity of primary importance is the entropy production 
rate (EPR) \cite{nicol,vel,espo}. 
The EPR vanishes for a closed system in the long-time limit 
that reaches a true TE. On the other hand, EPR 
is positive definite for a steady state that can emerge in an 
{\it open} system.  
The easiest way to model such a system in the context of 
chemical reactions is to assume that concentrations of some 
of the reacting species are held fixed \cite{Hqian1,Xie}. 
Under this 
condition, aptly known as the chemiostatic condition \cite{Min}, 
EPR tends to a non-zero constant, reflecting a steady 
dissipation rate (SDR) to sustain the system away from equilibrium 
\cite{HGepre2010}. 
The corresponding steady state is denoted as the 
non-equilibrium steady state (NESS) \cite{Hill1,HqjpcB,Jiang,GG2}. 
This concept has been extensively used in 
analyzing single-molecule kinetic experiments \cite{Xie,Min,Eng}. 
The NESS also includes the TE as a special case when 
detailed balance (DB) is obeyed \cite{Hqian2}, thus providing 
a very general framework.

Recently, an important progress was made in the theory 
and characterization of NESS, considering a master equation 
formalism \cite{Zia,Zia1,Zia2}. These studies have 
established that the classification of NESS 
requires {\it not only} the steady distribution 
(as in TE) but {\it also} the stationary fluxes or 
probability currents. This approach enables one to 
identify {\it all possible} combinations of transition rates that 
ultimately lead the system to the {\it same} NESS. 
However, these NESSs in general have different values of 
the EPR, and hence the SDR. 

This proposition prompts one to check 
(i) how states with the same EPR at NESS can be generated 
and  
(ii) whether there exist ways to distinguish these states. 
Here, we shall address both the issues by considering an 
enzyme-catalyzed reaction under chemiostatic 
condition. Expressing the EPR as a function of experimentally 
measurable reaction rate, 
we emphasize also that, the quantity that identifies 
the various NESSs having the same EPR is linked with the enzyme 
efficiency, a useful measure that is expressible in terms of 
enzyme kinetic constants.

\section{The system}

\begin{figure}[tbh]
\centering
\rotatebox{0}{
\includegraphics[width=7cm,keepaspectratio]{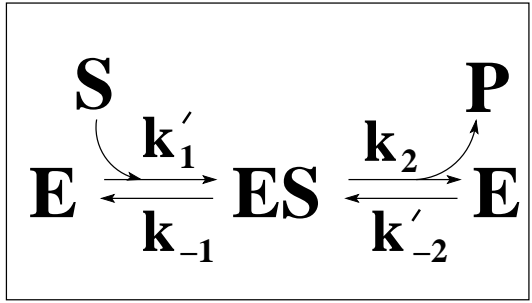}}
\caption{Schematic diagram of MM kinetics of 
enzyme catalysis with reversible product formation step 
under chemiostatic condition.}
\label{fig1}
\end{figure}

The basic scheme of enzyme catalysis within the Michaelis-Menten 
(MM) framework with reversible 
product formation step is shown in Fig.\ref{fig1}. 
Under chemiostatic condition, $[{\rm S}]$ and $[{\rm P}]$ 
are kept constant by 
continuous injection and withdrawal, respectively. 
This is the simplest model to mimic an open reaction 
system. 
Unlike the usual case of full enzyme recovery 
with total conversion of substrate into product in a 
closed system, 
here both the concentrations of free enzyme E and the enzyme-substrate 
complex ES reach a steady value. 
Also, instead of the rate of product formation, the progress of 
reaction is characterized by the rate of evolution of 
$[{\rm E}]$ (or $[{\rm ES}]$).

\subsection{Kinetics}

We define the pseudo-first-order rate constants 
as $k_1=k'_1[{\rm S}]$ and $k_{-2}=k'_{-2}[{\rm P}].$ 
Concentration of E is denoted by $c_1(t)$ and that 
of ES is given by $c_2(t).$ 
We have then
\begin{equation}
c_1(t)+c_2(t)=z.
\label{const}
\end{equation}
Here $z$ is a constant that stands for the total enzyme concentration. 
Then the rate of the reaction, $v(t)$, is written as 
\begin{equation}
v(t)=\dot{c_1}=-Kc_1(t)+(k_{-1}+k_2)z,
\label{reMM}
\end{equation}
where $K=(k_1+k_{-1}+k_2+k_{-2}).$ With the initial 
condition, $c_1(0)=z$, 
the time-dependent solution is given as 
\begin{equation}
c_1(t)=\frac{z}{K}\left((k_{-1}+k_2)+(k_1+k_{-2})e^{-Kt}\right).
\label{ct}
\end{equation}
The steady state enzyme concentration corresponds to the long-time 
limit of Eq.(\ref{ct}):
\begin{equation}
c_1^s=\left((k_{-1}+k_2)z\right)/K.
\label{cnsMM}
\end{equation}
At any steady state, we thus note
\begin{equation}
v(t)=\dot{c_1}\,=0\,=\,\dot{c_2}.
\label{vss}
\end{equation}

\subsection{Non-equilibrium thermodynamics}

The fluxes of the reaction system are defined pairwise as 
\cite{groot1,vel,espo}
\begin{equation}
J_1(t)=k_1c_1(t)-k_{-1}c_2(t),
\label{j1}
\end{equation}
\begin{equation}
J_2(t)=k_2c_2(t)-k_{-2}c_1(t).
\label{j2}
\end{equation}
From Eq.(\ref{const}), Eq.(\ref{reMM}), Eq.(\ref{j1}) 
and Eq.(\ref{j2}), one gets
\begin{equation}
\dot{c_1}=J_2(t)-J_1(t).
\label{rateflx}
\end{equation}
At the steady state, Eq.(\ref{rateflx}) leads to 
\begin{equation}
J_1^s=J_2^s=J^s.
\label{flxness}
\end{equation}
An NESS is characterized by a non-zero flux, $J^s\ne0$. 
At TE, the fluxes vanish for both 
the reactions. One may note, then the system satisfies DB.

The conjugate forces of the fluxes given in Eqs (\ref{j1})-(\ref{j2}) 
are defined as \cite{groot1}
\begin{equation}
X_1(t)=\mu_E+\mu_S-\mu_{ES}=T{\rm ln}\frac{k_1c_1(t)}{k_{-1}c_2(t)},
\label{f1}
\end{equation}
\begin{equation}
X_2(t)=\mu_{ES}-\mu_E-\mu_P=T{\rm ln}\frac{k_2c_2(t)}{k_{-2}c_1(t)}.
\label{f2}
\end{equation}
Corresponding to the scheme depicted in Fig.\ref{fig1}, 
the EPR is then given by \cite{denb,groot1}
\begin{equation}
\sigma(t)=\frac{1}{T}\sum_{i=1}^{2} J_i(t)X_i(t).
\label{eprMM}
\end{equation}
We set here (and henceforth) the Boltzmann constant $k_B=1$.
In the present case, the steady value of EPR becomes 
\begin{equation}
\sigma^s=\frac{1}{T}J^s(\mu_S-\mu_P).
\label{eprs}
\end{equation}
Therefore, unless the substrate and the product take part in equilibrium, 
the reaction system reaches an NESS with a SDR equal to $\sigma^s$.

\section{EPR close to NESS}

The problem is now transparent. If the rate constants 
become different, the steady concentrations will also differ. 
But, one can adjust them in such a way that $\sigma^s$ 
remains the same. In these situations, one 
needs an additional parameter to distinguish these states. 
To proceed, 
we define a small deviation in $c_1(t)$ around NESS as 
\begin{equation}
\delta=c_1(t)-c_1^s.
\label{delMM1}
\end{equation}
It then follows from Eq.(\ref{const}) that 
\begin{equation}
c_2(t)=c_2^s-\delta.
\label{delMM2}
\end{equation}
From Eq.(\ref{reMM}) and Eq.(\ref{delMM1}), the reaction rate 
becomes 
\begin{equation}
v(t)=-K\delta.
\label{vnsMM}
\end{equation}
Now, putting Eqs (\ref{j1})-(\ref{f2}) and 
Eqs (\ref{delMM1})-(\ref{vnsMM}) 
in Eq.(\ref{eprMM}) and taking only the first terms of 
the logarithmic parts, 
we obtain the EPR close to NESS as 
\begin{equation}
\sigma(t)={\rm A}_0+{\rm A}_1 v(t)+{\rm A}_2 v^2(t).
\label{eprMM1}
\end{equation}
Here 
\begin{equation}
{\rm A}_0=J^s{\rm ln}\frac{k_1k_2}{k_{-1}k_{-2}},
\label{x3}
\end{equation}
\begin{equation}
{\rm A}_1=-\frac{1}{K}\left((k_1+k_{-1}){\rm ln}\frac{k_1c_1^s}{k_{-1}c_2^s}
+(k_2+k_{-2}){\rm ln}\frac{k_{-2}c_1^s}{k_2c_2^s}\right),
\label{y3}
\end{equation}
\begin{equation}
{\rm A}_2=\frac{1}{K}\left(\frac{1}{c_1^s}+\frac{1}{c_2^s}\right).
\label{z3}
\end{equation}
As $v(t)$ vanishes at any steady state, the SDR at 
NESS is given by 
\begin{equation}
\sigma^s={\rm A}_0>0. 
\label{eprness}
\end{equation}
However, at TE, 
\begin{equation}
\sigma^s={\rm A}_0=0;
\label{epreq}
\end{equation}
one may check that here DB holds:
\begin{equation}
\frac{k_1k_2}{k_{-1}k_{-2}}=1.
\label{DB}
\end{equation}
Inspection of Eq.(\ref{eprMM1}) reveals that, 
near NESS, $\sigma(t)$ 
varies {\it linearly} with $v(t)$ with a slope ${\rm A}_1$. 
Thus, while ${\rm A}_0$ distinguishes an NESS from a 
true TE, ${\rm A}_1$ plays the same role in identifying systems 
with the {\it same} SDR but having {\it different} 
time profiles.

\section{Results and discussion}

In this section, we consider various situations 
where the reaction system reaches NESS with the same SDR. 
Focusing on Eq.(\ref{x3}), the different cases that 
keep ${\rm A}_0$ invariant are discussed next. 

\subsection{Variants with same SDR} 

\noindent
Case A: Any parent choice of rate constants. 

\noindent
Case B: Only $k_1$ and $k_2$ are exchanged.

\noindent
Case C: Only $k_{-1}$ and $k_{-2}$ are exchanged.

\noindent
Case D: Both $k_1,k_2$ and $k_{-1},k_{-2}$ are exchanged.

\noindent
Case E: Both $k_1,k_{-1}$ and $k_2,k_{-2}$ are exchanged.

\noindent
Case F: Both $k_1,k_{-2}$ and $k_2,k_{-1}$ are exchanged.

\begin{figure}[tbh]
\centering
\rotatebox{0}{
\includegraphics[width=7cm,keepaspectratio]{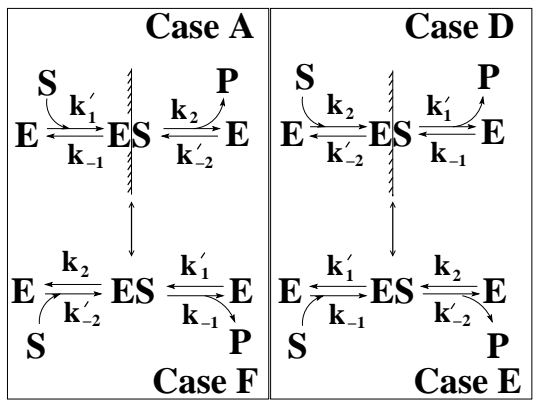}}
\caption{Schematic diagram showing the equivalence of 
the pairs A and F, and D and E, based on reflection symmetry.}
\label{fig2}
\end{figure}

\noindent
Case G: $k_1$ changed to $\alpha k_1$, $k_{-1}$ changed to 
$\alpha k_{-1}$, $k_2$ changed to $\beta k_2$ and 
$k_{-2}$ changed to $\beta k_{-2}$, 
such that $$\beta=\frac{1}{\alpha}=\frac{k_1+k_{-1}}{k_2+k_{-2}}.$$
It can be easily verified that cases D and E possess 
not only identical ${\rm A}_0$ but also the same ${\rm A}_1$ 
and ${\rm A}_2$. 
This is true for cases A and F as well. So, we do not 
consider cases E and F any further. A simple explanation 
of the equivalence is given in Fig.\ref{fig2} schematically, 
based on reflection symmetry. 

\subsection{Temporal profiles}

To explore the characteristics of various cases given above, 
we take the rate constants from the single molecule 
experimental study of English {\it et al.} \cite{Eng} on 
the {\it Escherichia coli} $\beta$-galactosidase enzyme. 
They are as follows: 
$k'_1=5.0$ E07 $\,{\rm M^{-1}s^{-1}},\,k_{-1}=1.83$ E04 ${\rm s^{-1}},\,
k_2=7.3$ E02 ${\rm s^{-1}}.$ 
We clarify that, in their study \cite{Eng}, $k_2$ had actually been 
shown to be a fluctuating quantity with a distribution. 
However, only an {\it average} value of $k_2$ 
will suffice our purpose. 
The constant substrate concentration is set at 
$[{\rm S}]=1.0$ E02 ${\rm \mu M}$ and thus, $k_1=k'_1[{\rm S}]
=5.0$ E03 ${\rm s^{-1}}$. 
We choose $k_{-2}=1.0$ E-05 ${\rm s^{-1}}$ to 
make the reaction scheme almost identical to the conventional 
MM kinetics. Here $\{k_i\}$ $(i=\pm1,\,\pm2)$ 
with magnitudes given above represents the 
parent choice of rate constants, {\it i.e.}, case A. 
The value of the constant 
$\beta=1/\alpha=3.2$ E01, in case G.

\begin{figure}[tbh]
\centering
\rotatebox{270}{
\includegraphics[width=7cm,keepaspectratio]{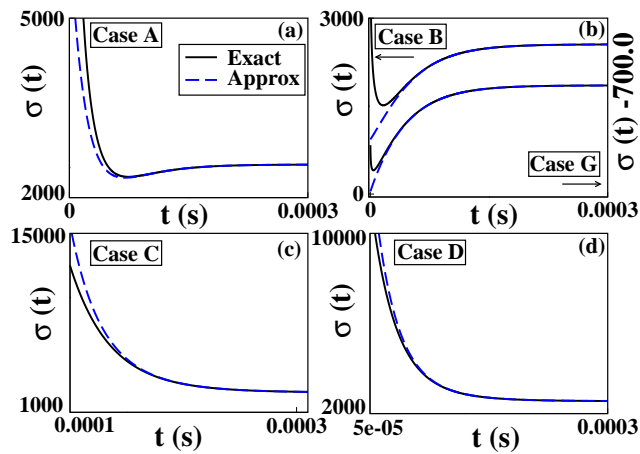}}
\caption{Evolution of EPR $\sigma(t)$ with time for 
various cases determined using the exact (Eq.(\ref{eprMM})) as well as 
the approximate (Eq.(\ref{eprMM1})) expressions. 
In panel (b), the EPR of case G 
is plotted as ($\sigma(t)-700.0$) for clarity.} 
\label{fig3}
\end{figure}

The time-evolution of EPR $\sigma(t)$, determined 
using both the exact (Eq.(\ref{eprMM})) 
and the approximate (Eq.(\ref{eprMM1})) expressions, are shown 
in Fig.\ref{fig3}, for the various cases. 
The concentrations $c_1,\,c_2$ are made 
dimensionless by scaling with respect to the total enzyme 
concentration $z$. This ensures that $\sigma(t)$ has the 
unit of ${\rm s^{-1}}$. 
From the figure, it is evident that Eq.(\ref{eprMM1}) 
nicely approximates the behavior near NESS. 
Specifically, the curves of exact and approximate 
cases merge quite well for any $t\ge\,1.5$ E-04 s. 
\begin{figure}[tbh]
\centering
\rotatebox{270}{
\includegraphics[width=7cm,keepaspectratio]{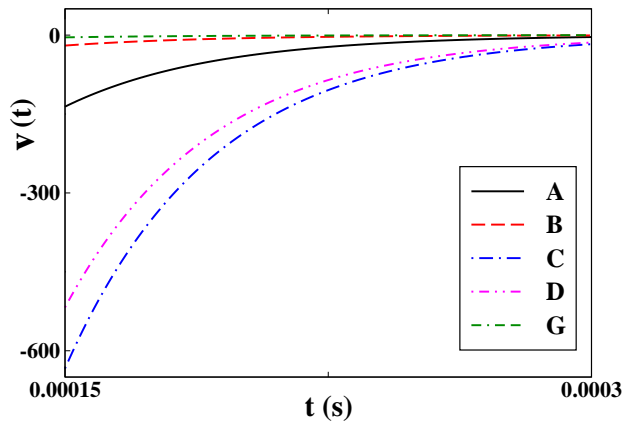}}
\caption{Variation of reaction rate $v(t)$ as a function 
of time for different cases indicated in the plot.} 
\label{fig4}
\end{figure}

\begin{figure}[tbh]
\centering
\rotatebox{270}{
\includegraphics[width=7cm,keepaspectratio]{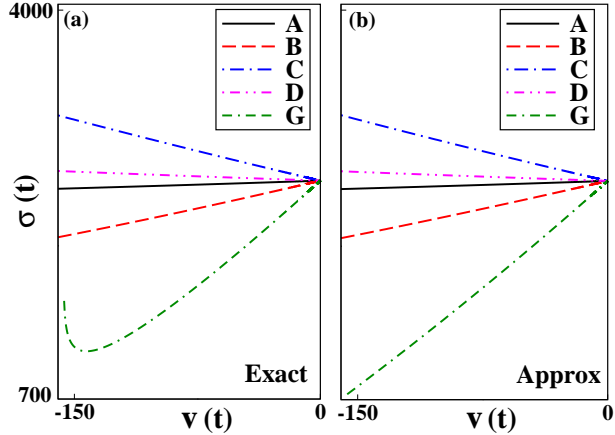}}
\caption{Variation of EPR $\sigma(t)$ as a function of 
reaction rate $v(t)$ for different cases indicated in the plot 
using (a) exact (Eq.(\ref{eprMM})) and (b) approximate 
(Eq.(\ref{eprMM1})) expressions.}
\label{fig5}
\end{figure}

The evolution of reaction rate $v(t)$ is shown in 
Fig.\ref{fig4} for all the distinct cases. The curves are 
displayed over a time-span where Eq.(\ref{eprMM1}) is 
valid, as mentioned above. This gives us a quantitative 
understanding of the magnitude of $v(t)$ up to 
which the {\it close to} NESS approximation, and 
hence Eq.(\ref{eprMM1}), is valid. 
We note the variation of $\sigma(t)$ as a function of 
$v(t)$ in all the relevant cases in Fig.\ref{fig5}. 
Both the exact (Fig.\ref{fig5}(a)) as well as the approximate 
results (Fig.\ref{fig5}(b)) are shown. 
Two features are interesting. 
First, in all the situations, 
the system reaches an NESS with identical 
$\sigma^s={\rm A}_0=2.553$ E03 ${\rm s^{-1}}$. 
Secondly, the quantity that distinguishes one case 
from the other is the slope ${\rm A}_1$ of $\sigma(t)$ vs. $v(t)$ 
curve near the NESS. This slope can be positive as well as 
negative. 

\subsection{Total entropy production and enzyme efficiency}

One may like to next investigate the role of the 
rate constants in governing the overall dissipation in various cases. 
Specifically, we like to enquire if the efficiency 
of the enzyme has anything to do with the 
total dissipation. In this context, it may be recalled 
that, the conventional MM kinetics requires the rate constant 
$k_{-2}$ to be negligible compared with the others. 
So, the enzyme kinetic constants, like the MM constant 
${\rm K_M}=\frac{k_{-1}+k_2}{k'_1}$ and 
catalytic efficiency $\eta =k_2/{\rm K_M}$, are meaningful 
in the limit $k_{-2}\to 0$. 
Our choice of parent rate constants ensures that in case A, 
the system follows MM kinetics. 
Case B, which leaves $k_{-2}$ unchanged and case G, 
which changes $k_{-2}$ to $\beta k_{-2}$ (with $\beta=3.2$ E01), 
can also be included within the MM scheme. 
But, cases C to F, which exchange $k_{-2}$ with any one of the 
other bigger rate constants, can not follow the usual 
MM kinetics. Therefore, we focus on cases A,B and G 
in finding any possible connection between 
the kinetic constants of the enzyme and the total dissipation. 
While the SDR $\sigma^s$ is the same 
for all of them, the 
time-integrated EPR, giving the total entropy 
production, is different. We define it as 
\begin{equation}
S_I=\int_{0}^{\tau}\sigma(t)dt.
\label{eprint}
\end{equation}
The upper limit $\tau$ 
is fixed at such a time when all the systems reach NESS. 
In the present set of cases, we find that setting $\tau=1.0$ E-03 s 
is satisfactory. 
The values of ${\rm K_M},\,\eta$ and $S_I$ (determined 
by integrating $\sigma(t)$ from Eq.(\ref{eprMM})) 
are listed in Table \ref{tab1}, along with the slope ${\rm A}_1$ 
[see Eq.(\ref{eprMM1})]. 
It is clear from the data that, in going 
from case A to case G, ${\rm K_M}$ gradually increases, whereas 
$\eta$ falls. Both these features indicate that the 
enzyme becomes {\it less efficient}. 
More interesting is to note that the corresponding $S_I$ values 
also exhibit a decreasing trend from case A to case G. 
Thus, we can say that, with identical SDR, 
the more efficient enzyme 
(bigger $\eta$ and smaller ${\rm K_M}$) involves higher {\it total} 
dissipation. This can be rationalized 
by the fact that, higher efficiency corresponds to a 
{\it faster} conversion 
of substrate into product. This implies an increased 
irreversibility in the process. Consequently, a higher entropy 
production is noted.

\begin{table}[tbh]
\caption{Values of the quantities ${\rm A}_1$, ${\rm K_M}$, 
$\eta$ and $S_I$ for cases A,B and G.}
\begin{center}
\begin{tabular}{lcccc}
\hline
\hline
Case & ${\rm A}_1$ & ${\rm K_M}$ & $\eta$ & $S_I$ \\
\hline
A & 4.715 E-01 & 3.806 E-04 & 1.918 E06 & 2.68 E0 \\
B & 3.256 E0 & 3.192 E-03 & 1.566 E06 & 2.48 E0 \\
G & 1.257 E01 & 1.524 E-02 & 1.529 E06 & 2.47 E0 \\
\hline
\hline
\end{tabular}
\end{center}
\label{tab1}
\end{table} 

Before ending this section, we mention briefly the fate of the different 
situations when DB, Eq.(\ref{DB}), gets satisfied. 
In this scenario, whatever 
be the values of the rate constants, the final EPR is trivially zero as 
the reaction system reaches TE [see Eq.(\ref{epreq})]. 
For the same reason, ${\rm A}_1$ also becomes zero 
[see Eq.(\ref{y3})]. 
However, it follows from Eq.(\ref{eprMM1}) that, EPR 
varies {\it quadratically} 
with $v(t)$ near TE. Then, in principle, ${\rm A}_2$ {\it can} 
distinguish systems reaching TE. It is easy to see 
from Eq.(\ref{z3}) that, 
cases A, B and G possess different values for 
${\rm A}_2$ and hence they can be identified by following the 
behavior of EPR with the reaction rate.

\section{Extension to general reaction systems}

\begin{figure}[tbh]
\centering
\rotatebox{0}{
\includegraphics[width=7cm,keepaspectratio]{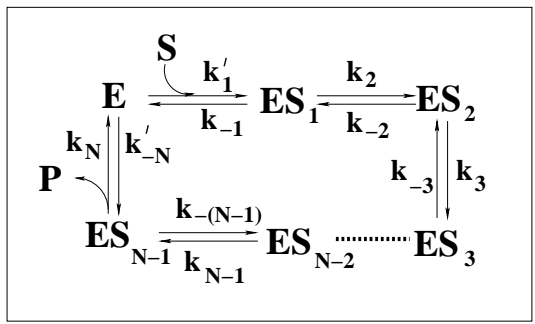}}
\caption{Schematic diagram of enzyme kinetics with (N-1) number 
of intermediates under chemiostatic condition.}
\label{fig6}
\end{figure}

The MM kinetics, shown in Fig.\ref{fig1}, with a single 
intermediate in the form of the ES complex, is {\it exactly} solvable. 
We now generalize this 
scheme to an enzyme catalysis reaction having N number of 
species. 
These include the free enzyme E and 
(N-1) intermediates, 
under similar chemiostatic condition as discussed in Section II. 
The reaction scheme is depicted in Fig.\ref{fig6}. 
Essentially, the species ${\rm ES_j},\,
(j=1,\cdots,N-1)$ refer to the various conformers of the 
enzyme-substrate complex. 
The corresponding rate equations are given as 
\begin{equation}
\dot{c_i}=-(k_i+k_{-(i-1)})c_i(t)+k_{i-1}c_{i-1}(t)+k_{-i}c_{i+1}(t),
\label{dai}
\end{equation}
with $c_i(t) \,(i=1,\cdots,N)$ being the concentration of 
species ${\rm ES}_{(i-1)}$ at time $t$. 
The following periodic boundary conditions hold:
$$i-1=N,\, {\rm for}\, i=1,$$
$$i+1=1,\, {\rm for}\, i=N.$$
We have set $k_1=k'_1[{\rm S}]$ and $k_{-N}=k'_{-N}[{\rm P}].$ 
The flux $J_i$ due to the i-th reaction is defined as
\begin{equation}
J_i(t)=k_i c_i(t)-k_{-i}c_{i+1}(t).
\label{Ji}
\end{equation}
The expression of EPR then becomes 
\begin{equation}
\sigma_N(t)=\sum_{i=1}^{N}J_i(t){\rm ln}\frac{k_ic_i(t)}{k_{-i}c_{i+1}(t)}.
\label{eprc}
\end{equation}

\subsection{EPR as a functional of reaction rate near NESS}

It is generally not possible to solve the set of 
coupled equations analytically for a system of arbitrary 
size. However, again focusing on a situation close to the NESS, 
one can get some insights. For that purpose, 
we define small deviations in species concentrations from their 
respective NESS values as 
\begin{equation}
\delta_i(t)=c_i(t)-c_i^s. 
\label{del}
\end{equation} 
For a short time interval $\tau$, using finite difference approximation, 
one gets 
\begin{equation}
\dot{\delta_i}=\dot{c_i} \approx \delta_i/\tau. 
\label{del1}
\end{equation}
Putting Eqs (\ref{del})-(\ref{del1}) in Eq.(\ref{dai}), we get 
\begin{equation}
\left(1-(k_i+k_{-(i-1)})\tau\right)\delta_i(t)+
k_{i-1}\tau\delta_{i-1}(t)+k_{-i}\tau\delta_{i+1}(t)=0.
\label{delness}
\end{equation}
As the reactions are coupled, so the $\delta_i$s are related 
to each other and can be expressed in terms of any 
one of them, say $\delta_1.$ Then, one can write 
\begin{equation}
\delta_i=f_i\delta_1, \quad {\rm with}\,\, f_1=1.
\label{relatn}
\end{equation}
Next we will discuss the scheme to determine the $f_i$s. 

The set of coupled equations (\ref{delness}), with the help 
of Eq.(\ref{relatn}), can be cast in the matrix form
\begin{equation}
{\bf Mf=0}.
\label{mat}
\end{equation}
Here ${\bf f}$ is a $N\times1$ matrix with ${\bf f^T}=
\left(f_1, f_2,\cdots,f_N\right)$ and ${\bf M}$ is a 
$N\times N$ matrix with the property
$$M_{ij}\ne0,\, {\rm for}\, j=i,i-1,i+1$$
\begin{equation}
M_{ij}=0, \, {\rm otherwise}.
\label{matel}
\end{equation}
The non-zero matrix elements are 
\begin{equation}
M_{ii}=\left(1-(k_i+k_{-(i-1)})\tau\right),
\label{matel1}
\end{equation}
\begin{equation}
M_{i,i-1}=k_{i-1}\tau,
\label{matel2}
\end{equation}
\begin{equation}
M_{i,i+1}=k_{-i}\tau.
\label{matel3}
\end{equation}
From Eq.(\ref{mat}) and Eq.(\ref{matel}), we obtain a 
recursion relation 
\begin{equation}
f_j=-\frac{(M_{j-1,j-2}f_{j-2}+M_{j-1,j-1}f_{j-1})}{M_{j-1,j}},\, 
{j=2,3,\cdots,N},
\label{recur}
\end{equation}
with the boundary conditions:
$$f_0=f_N,\,\,M_{j,0}=M_{j,N}.$$
The first of the relations becomes 
\begin{equation}
f_2=-\frac{M_{1N}f_N+M_{11}}{M_{12}}.
\label{recur1}
\end{equation}
Then, it is easy to follow from Eq.(\ref{recur}) that, all the other 
$f_j$s can be expressed in terms of $f_N.$ 
From the condition 
\begin{equation}
\sum_{i=1}^{N}c_i={\rm constant},
\end{equation}
we get 
\begin{equation}
\sum_{i=1}^{N}\delta_i=0,
\end{equation}
and using Eq.(\ref{relatn}), we have 
\begin{equation}
\sum_{i=2}^{N}f_i=-1.
\label{sumf}
\end{equation}
From Eqs (\ref{matel1})-(\ref{recur1}) and Eq.(\ref{sumf}), one 
can determine the $f_i$s in Eq.(\ref{relatn}).

We are now ready to explore the EPR near the NESS. 
From Eq.(\ref{dai}), we have
\begin{equation}
J_i^s=J^s,\quad (i=1,\cdots,N)
\label{jns}
\end{equation}
at NESS. As we have chosen to express all the 
deviations in concentration from the NESS in terms of $\delta_1$, 
so we take the reaction rate as $v(t)=\dot{a_1}$. 
Then, from Eq.(\ref{dai}) with $i=1$ and using Eq.(\ref{relatn}) 
along with the periodic boundary 
conditions, we get near NESS 
\begin{equation}
v(t)=R\delta_1,
\label{vns}
\end{equation}
where
\begin{equation}
R=-(k_1+k_{-N})+k_{N}f_N+k_{-1}f_2.
\label{r1}
\end{equation}
Now putting Eq.(\ref{del}), Eq.(\ref{relatn}), 
Eq.(\ref{jns}) and Eq.(\ref{vns}) in Eq.(\ref{eprc}) and 
also using the smallness of $\delta_i$s, the 
EPR near NESS becomes
$$\sigma_N(t)=\sum_{i=1}^{N}\left(J^s+k_i\delta_i-k_{-i}\delta_{i+1}\right)
\left({\rm ln}\frac{k_ic_i^s}{k_{-i}c_{i+1}^s}+
\frac{\delta_i}{c_i^s}-\frac{\delta_{i+1}}{c_{i+1}^s}\right)$$
\begin{equation}
\sigma_N(t)={\rm A}_N^{(0)}+{\rm A}_N^{(1)} v(t)+{\rm A}_N^{(2)} v^2(t),
\label{epr1}
\end{equation}
with 
\begin{equation}
{\rm A}_N^{(0)}=J^s{\rm ln}\frac{\prod_{i=1}^Nk_i}{\prod_{i=1}^Nk_{-i}},
\label{x1}
\end{equation}
\begin{equation}
{\rm A}_N^{(1)}=\frac{1}{R}\sum_{i=1}^N \left(k_if_i-k_{-i}f_{i+1}\right)
{\rm ln}\frac{k_ic_i^s}{k_{-i}c_{i+1}^s},
\label{y1}
\end{equation}
\begin{equation}
{\rm A}_N^{(2)}=\frac{1}{R^2}\sum_{i=1}^N \left(k_if_i-k_{-i}f_{i+1}\right)
\left(\frac{f_i}{c_i^s}-\frac{f_{i+1}}{c_{i+1}^s}\right).
\label{z1}
\end{equation}
Eq.(\ref{epr1}) is the generalized version of Eq.(\ref{eprMM1}), 
confirming that expression of the EPR as a functional of reaction 
rate possesses a universal character. 

\subsection{Cases with invariant SDR}

The next task is, whether states having the same SDR, 
{\it i.e.}, identical ${\rm A}_N^{(0)}$, 
can be generated for the N-cycle. An obvious clue comes 
from the invariance of 
a cycle under rotation. Thus, if the steady concentrations 
$c_i^s$ are represented as N points uniformly placed on a circle, 
then rotations by an angle $\theta$, defined as 
\begin{equation}
\theta=\frac{2\pi j}{N},\quad j=1,\cdots,(N-1),
\label{ang}
\end{equation}
will just redistribute the $c_i^s$ values. This keeps the steady 
flux $J^s$ in Eq.(\ref{x1}) unchanged. Therefore, for a 
N-cycle, there are {\it at least} (N-1) ways 
to interchange the rate constants $k_{\pm i}$ that will lead the 
reaction system to states with the same SDR. 
We illustrate this result here by taking the simplest 
non-trivial case of a triangular network as an example. 

One can see from Eq.(\ref{ang}) that, for a triangular 
network with $N=3$, {\it at least} 
two kinds of changes of the rate constants keep the SDR unchanged.  
They are given below:

\noindent
Case 1. Any parent choice of rate constants. 

\noindent
Case 2. Change $k_{\pm i}\to k_{\pm (i+1)},\, (i=1,\cdots,N)$ 
with the boundary condition $k_{\pm (N+1)}=k_{\pm 1}$.  

\noindent
Case 3. Change $k_{\pm 1}\to k_{\pm 3}$, $k_{\pm 2}\to k_{\pm 1}$ and 
$k_{\pm 3}\to k_{\pm 2}$. 

One can generate additional ways to keep ${\rm A}_N^{(0)}$ 
fixed with some added constraints on the rate constants. 
Two pairs of situations [cases 4 and 5, and 6 and 7] 
are the following:

\noindent
Case 4. Any parent choice of rate constants with $k_1=k_{-1}$. 

\noindent
Case 5. Change $k_1\to k_2$, $k_2\to k_3$, $k_3\to k_{-1}$, 
$k_{-1}\to k_{-2}$, $k_{-2}\to k_{-3}$ and $k_{-3}\to k_{1}$. 

\noindent
Case 6. Any parent choice of rate constants with $k_3=k_{-3}$.

\begin{figure}[tbh]
\centering
\rotatebox{270}{
\includegraphics[width=8cm,keepaspectratio]{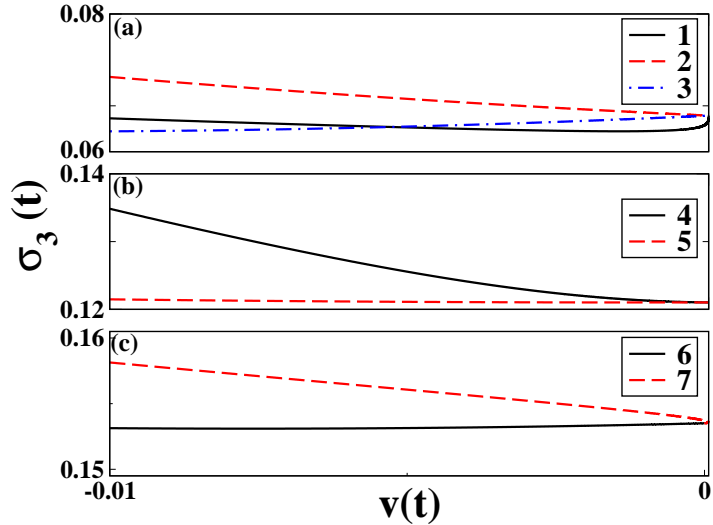}}
\caption{Variation of EPR $\sigma_3(t)$ (Eq.(\ref{eprc})) 
as a function of reaction rate $v(t)$ for 
cases (a) 1,2 and 3, (b) 4 and 5, (c) 6 and 7.}
\label{fig7}
\end{figure}

\noindent
Case 7. Change $k_1\to k_{-3}$, $k_{-3}\to k_{-2}$, $k_{-2}\to k_{-1}$, 
$k_{-1}\to k_{3}$, $k_{3}\to k_{2}$ and $k_{2}\to k_{1}$.\\
All the above variants have been numerically studied and shown in 
Fig.\ref{fig7} where the EPR, determined exactly by 
Eq.(\ref{eprc}), is plotted as a function of reaction rate 
$v(t)=\dot{a_1}$ for each of the cases. 
It is evident from the figure that the SDR are identical for the 
respective bunch of cases. 
But they can be distinguished by following the 
$\sigma_3(t)$ vs. $v(t)$ curve in the small-$v(t)$ regime.

\section{Conclusion}

In summary, the present endeavor has been to characterize 
steady states with the same non-zero SDR. We have found that the 
variation of EPR with the reaction rate 
near completion of the reaction is a nice indicator 
to distinguish such states. 
Particularly important is the role of the slope 
of $\sigma(t)$ vs. $v(t)$ curve near $v(t)=0$. 
This has been substantiated by studying 
enzyme-catalysed reactions as an exactly-solvable 
test case. 
We have also noticed, the leading term that accounts for 
the variation depends on the rate constants, 
more specifically on the enzyme efficiency. It is 
gratifying to observe that the more efficient enzyme 
incurs higher total dissipation. The physical appeal 
is immediate. A more efficient enzyme 
approaches the steady state more quickly. This implies 
the process becomes more irreversible. 
Hence, $S_I$ becomes higher. One more notable point 
is the following. The SDR is equal to the steady heat 
dissipation rate. Our study reveals that enzymes 
with very different efficiencies can show the same heat 
dissipation rate at steady state. 
An extension to cases of 
higher complexities involving various conformers of 
the enzyme-substrate complex has also been envisaged. 
Further studies along 
this line on enzymes with multiple sites may be worthwhile.

\section*{Acknowledgment}

K. Banerjee acknowledges the University Grants Commission (UGC), India 
for Dr. D. S. Kothari Fellowship. K. Bhattacharyya thanks CRNN, CU, 
for partial financial support.

\end{document}